\documentclass{llncs}
\usepackage{version}
\usepackage{pcl}

\title{A Survey of Paraconsistent Logics}
\author{C. A. Middelburg}
\institute{Informatics Institute, Faculty of Science,
           University of Amsterdam, \\
           Science Park~107, 1098~XG Amsterdam, the Netherlands \\
           \email{C.A.Middelburg@uva.nl}}

\begin{document}
\maketitle

\begin{abstract}
A survey of paraconsistent logics that are prominent representatives of
the different approaches that have been followed to develop
paraconsistent logics is provided.
The paraconsistent logics that will be discussed are an enrichment of
Priest's logic LP, the logic RM$_3$ from the school of relevance logic,
da Costa's logics C$_n$, Ja\'{s}kowski's logic D$_2$, and Subrahmanian's
logics P$\tau$.
A deontic logic based on the first of these logics will be discussed as
well.
Moreover, some proposed adaptations of the AGM theory of belief revision
to paraconsistent logics will be mentioned.
\end{abstract}

\section{Introduction}
\label{sect-introduction}

Paraconsistent logics are those logics that do not have the property
that any formula can be deduced from every set of hypotheses that
contains contradictory formulas.
The paraconsistent logics that have been proposed differ in many ways.
The differences are mostly minor, but occasionally major.
Whether one paraconsistent logic is more plausible than another is
fairly difficult to make out.

A logic with the property that any formula can be deduced from every set
of hypotheses that contains contradictory formulas but one is far from a
reasonable paraconsistent logic.
Such a logic is in a certain sense a minimal paraconsistent logic.
In a quest for paraconsistent logics that are maximally paraconsistent,
many different paraconsistent logics have been proposed.
Some of them are in fact maximally paraconsistent in a well-defined
sense.
However, there are other properties than maximal paraconsistency that
are usually considered characteristic of reasonable paraconsistent
logics.
Among them is the property that the logic concerned does not validate
deductions forbidden by classical logic that are not essential for
paraconsistency.
There are also less technical properties that are sometimes considered
important.
Among them are the ease with which the axiom schemas and inference rules
of the logic concerned can be memorized and the ease with which the
semantics of the logic concerned can be memorized.

As a rule, survey articles and handbook chapters on the subject of
paraconsistent logic concentrate on explaining the nature of and
motivation for the subject, giving a history of the subject, and/or
surveying the basic techniques used to develop paraconsistent logics
(see e.g.~\cite{Pri84a,Pri02a}).
Individual paraconsistent logics proposed in the scientific literature
are only touched to illustrate the techniques explained.
The exceptions where a survey of several paraconsistent logics is
provided, concentrate on discussing logics that have been developed
following a particular approach (see e.g.~\cite{Arr80a,dKB07a}).
Gaining a basic understanding of a number of different paraconsistent
logics and their interrelationships still requires an extensive study of
scientific publications which contain many theoretical details that are
not relevant to a basic understanding.

In this note, a survey of paraconsistent logics that are prominent
representatives of the different approaches that have been followed to
develop paraconsistent logics is provided.
The survey is made so as to allow for gaining a basic understanding of
the logics in question and their interrelationships.
For each approach, the logic that has been selected as the prominent
representative is one of the representatives for which the number of
publications about work on it and/or the number of publications that
cite the publications about work on it is relatively high.
In a strict sense, the survey covers only a rather narrow group of
paraconsistent logics.
However, it is not so narrow as it seems at first, because the selected
representative for each approach is in general closely related to most
other representatives of the approach concerned.

The different approaches that have been followed to develop
paraconsistent logics are:
\begin{itemize}
\item
the \emph{three-valued approach}:
classical logic is turned into a logic based on three truth values:
true, false and both-true-and-false;
\item
the \emph{relevance approach}:
classical logic is adapted to the idea that the antecedent of an
implication must be relevant to its consequent;
\item
the \emph{non-truth-functional approach}:
classical logic is turned into a logic based on a non-truth-functional
version of negation;
\item
the \emph{non-adjunctive approach}:
classical logic is adapted to the idea that the inference of $A \And B$
from $A$ and $B$ must fail;
\item
the \emph{annotation approach}:
classical logic is changed into a logic where atomic formulas are
annotated with believed truth values.
\end{itemize}
The representative of these approaches that will be discussed in this
note are
LP$^{\IImpl}$, which is Priest's logic LP~\cite{Pri79a} enriched with an
implication connective for which the standard deduction theorem holds,
the logic RM$_3$~\cite{AB75a} from the school of relevance logic,
da Costa's logics C$_n$~\cite{daC74a},
Ja\'{s}kowski's logic D$_2$~\cite{Jas99a,Jas99b}, and
Subrahmanian's logics P$\tau$~\cite{dSV91a}, respectively.

Only propositional logics will be discussed.
Details about the extensions of LP$^{\IImpl}$, RM$_3$, C$_n$ and P$\tau$
to the corresponding predicate logics can be found in~\cite{Pri02a},
\cite{DR02a}, \cite{dKB07a} and~\cite{ANA07a}, respectively.%
\footnote
{In~\cite{dKB07a}, reference is made to several publications in which a
 formulation of D$_2$ can be found.
 All formulations concerned, as well as the one in~\cite{Kd79a}, are
 incorrect (cf.~\cite{Ciu08a}).
 Consequently, the formulations of the extension of D$_2$ to a predicate
 logic that can be found in some of these publications are incorrect as
 well.
}
These extensions are exactly as to be expected if universal
quantification and existential quantification are regarded as
generalized conjunction and generalized disjunction, respectively.
Therefore, their discussion adds little to a basic understanding.

The survey of paraconsistent logics provided in this note stems from
interest in modelling legal reasoning.
However, legal reasoning is not only reasoning in the presence of
inconsistent information, but also reasoning concerned with normative
expressions such as obligatory, permissible, and prohibited.
Therefore, a deontic logic based on one of the paraconsistent logics
discussed in the survey, LP$^{\IImpl}$, is discussed in this note as
well.
Moreover, it is reasonable to assume that legal reasoning is done by
rational persons who revise their beliefs in the light of new
information.
For that reason, proposed adaptations of the main theory of belief
revision, the AGM theory, to paraconsistent logics and other work
relevant to belief revision in the presence of inconsistent beliefs are
also mentioned.

\section{Preliminaries}
\label{sect-preliminaries}

Each logic that will be discussed in this note has the following logical
connectives:
an implication connective $\IImpl$,
a conjunction connective $\And$,
a disjunction connective $\Or$, and
a negation connective~$\Not$.%
\footnote
{Each of the connectives is requisite in the formulation of a
 representative of at least one of the approaches mentioned in
 Section~\ref{sect-introduction}.
}
Bi-implication is in all cases defined as an abbreviation:
$A \IIff B$ stands for $(A \IImpl B) \And (B \IImpl A)$.

In LP$^{\IImpl}$, RM$_3$, C$_n$ and D$_2$, the formation rules for
formulas are the same as in classical propositional logic.
Hence, the languages of LP$^{\IImpl}$, RM$_3$, C$_n$ and D$_2$ are
identical.
In P$\tau$, the formation rules for formulas differ in that the atomic
formulas are annotated propositional variables instead of propositional
variables.

For each logic that will be discussed in this note, a Hilbert-style
formulation will be given.
In those formulations, $A$, $B$ and $C$ will be used as meta-variables
ranging over all formulas of the logic concerned.

In the case of a Hilbert-style formulation, a proof of a formula $A$
from a set of formulas $\Gamma$ in a logic $\mathcal{L}$ is a sequence 
of formulasending with $A$ such that each formula in the sequence is 
either an axiom, or a formula in $\Gamma$, or a formula that follows 
from previous formulas in the sequence by one of the rules of inference.

The logical consequence relation of a logic $\mathcal{L}$, denoted by 
$\Ent{\mathcal{L}}$, is the binary relation between sets of formulas 
and formulas defined as follows: $\Gamma \Ent{\mathcal{L}} A$ iff there 
exists a proof of $A$ from $\Gamma$ in $\mathcal{L}$.

A logic $\mathcal{L}$ is called a \emph{paraconsistent} logic if its 
logical consequence relation $\Ent{\mathcal{L}}$ satisfies the condition 
that there exist a set $\Gamma$ of formulas of $\mathcal{L}$ and 
formulas $A$ and $B$ of $\mathcal{L}$ such that 
$\Gamma \Ent{\mathcal{L}} A$ and $\Gamma \Ent{\mathcal{L}} \Not A$, but 
not $\Gamma \Ent{\mathcal{L}} B$.

\section{Priest's Paraconsistent Logic LP$^{\IImpl}$}
\label{sect-LPiimpl}

In~\cite{Pri79a}, Priest proposes the paraconsistent propositional logic
LP (Logic of Paradox).
The logic LP$^{\IImpl}$ introduced in this section is LP enriched with
an implication connective for which the standard deduction theorem
holds.
This logic is also known under the following names: PAC~\cite{Avr91a},
PI$^s$~\cite{Bat80a} and pure CLuNs~\cite{BC04a}.
The fragment without the implication connective was already suggested by
Asenjo in 1966 (see~\cite{Ase66a}).

A Hilbert-style formulation of LP$^{\IImpl}$ is given in
Table~\ref{proofsystem-LPiimpl}.
\begin{table}[!ht]
\caption{Hilbert-style formulation of LP$^{\IImpl}$}
\label{proofsystem-LPiimpl}
\begin{eqntbl}
\begin{axcol}
\mathbf{Axiom\; Schemas:}
\\
A \IImpl (B \IImpl A)
\\
(A \IImpl (B \IImpl C)) \IImpl ((A \IImpl B) \IImpl (A \IImpl C))
\\
((A \IImpl B) \IImpl A) \IImpl A
\\
(A \And B) \IImpl A
\\
(A \And B) \IImpl B
\\
A \IImpl (B \IImpl (A \And B))
\\
A \IImpl (A \Or B)
\\
B \IImpl (A \Or B)
\\
(A \IImpl C) \IImpl ((B \IImpl C) \IImpl ((A \Or B) \IImpl C))
\eqnsep
\mathbf{Rule\; of\; Inference:}
\\
\Infrule{A \quad A \IImpl B}{B}
\end{axcol}
\qquad
\begin{axcol}
{}
\\
\Not \Not A \IIff A
\\
\Not (A \IImpl B) \IIff A \And \Not B
\\
\Not (A \And B) \IIff \Not A \Or \Not B
\\
\Not (A \Or B) \IIff \Not A \And \Not B
\\
{}
\\
A \Or \Not A
\end{axcol}
\end{eqntbl}
\end{table}
In this formulation, which is taken from~\cite{Avr91a}, $A$, $B$, and 
$C$ are used as meta-variables ranging over all formulas of
LP$^{\IImpl,\False}$.
The axiom schemas on the left-hand side of
Table~\ref{proofsystem-LPiimpl} and the single inference rule (modus
ponens) constitute a Hilbert-style formulation of the positive fragment
of classical propositional logic.
The first four axiom schemas on the right-hand side of
Table~\ref{proofsystem-LPiimpl} allow for the negation connective to be
moved inward.
We get a sound and complete formulation of the propositional part of the
paraconsistent logic N$^-$, which was proposed by Nelson
in~\cite{Nel59a}, if we add these axiom schemas to a Hilbert-style
formulation of the positive fragment of intuitionistic propositional
logic.
The fifth axiom schema on the right-hand side of
Table~\ref{proofsystem-LPiimpl} is the law of the excluded middle.
This axiom schema can be thought of as saying that, for every
proposition, the proposition or its negation is true, while leaving open
the possibility that both are true.
If we add the axiom schema $\Not A \IImpl (A \IImpl B)$, which says that
any proposition follows from a contradiction, to the given Hilbert-style
formulation of LP$^{\IImpl}$, then we get a Hilbert-style formulation of
classical propositional logic (see e.g.~\cite{Avr91a}).

We use the symbol $\pEnt$ to denote the syntactic logical consequence 
relation induced by the axiom schemas and inference rule of 
LP$^{\IImpl,\False}$.

The following outline of the semantics of LP$^{\IImpl}$ is based
on~\cite{Avr91a}.
Like in the case of classical propositional logic, meanings are assigned
to the formulas of LP$^{\IImpl}$ by means of valuations.
However, in addition to the two classical truth values $\true$ (true)
and $\false$ (false), a third meaning $\both$ (both true and false) may
be assigned.

A \emph{valuation} for LP$^{\IImpl}$ is a function $\nu$ from the set of
all formulas of LP$^{\IImpl}$ to the set $\{\true,\false,\both\}$ such
that for all formulas $A$ and $B$ of LP$^{\IImpl}$:
\pagebreak[2]
\begin{eqnarray*}
\val{A \IImpl B}{\nu} & = &
 \left \{
 \begin{array}{l@{\;\;}l}
 \true           & \mathrm{if}\; \val{A}{\nu} = \false \\
 \val{B}{\nu} & \mathrm{otherwise},
 \end{array}
 \right.
\\
\val{A \And B}{\nu} & = &
 \left \{
 \begin{array}{l@{\;\;}l}
 \true  & \mathrm{if}\; \val{A}{\nu} = \true  \;\mathrm{and}\;
                        \val{B}{\nu} = \true  \\
 \false & \mathrm{if}\; \val{A}{\nu} = \false \;\mathrm{or}\;
                        \val{B}{\nu} = \false \\
 \both  & \mathrm{otherwise},
 \end{array}
 \right.
\\
\val{A \Or B}{\nu} & = &
 \left \{
 \begin{array}{l@{\;\;}l}
 \true  & \mathrm{if}\; \val{A}{\nu} = \true  \;\mathrm{or}\;
                        \val{B}{\nu} = \true  \\
 \false & \mathrm{if}\; \val{A}{\nu} = \false \;\mathrm{and}\;
                        \val{B}{\nu} = \false \\
 \both  & \mathrm{otherwise},
 \end{array}
 \right.
\\
\val{\Not A}{\nu} & = &
 \left \{
 \begin{array}{l@{\;\;}l}
 \true  & \mathrm{if}\; \val{A}{\nu} = \false \\
 \false & \mathrm{if}\; \val{A}{\nu} = \true \\
 \both  & \mathrm{otherwise}.
 \end{array}
 \right.
\end{eqnarray*}
The classical truth-conditions and falsehood-conditions for the logical
connectives are retained.
Except for implications, a formula is classified as both-true-and-false
exactly when when it cannot be classified as true or false by the
classical truth-conditions and falsehood-conditions.
The definition of a valuation given above shows that the logical 
connectives of LP$^{\IImpl,\False}$ are (three-valued) truth-functional, 
which means that each $n$-ary connective represents a function from 
$\{\true,\false,\both\}^n$ to $\{\true,\false,\both\}$.

For LP$^{\IImpl,\False}$, the semantic logical consequence relation, 
denoted by $\mEnt$, is based on the idea that a valuation $\nu$ 
satisfies a formula $A$ if $\val{A}{\nu} \in \{\true,\both\}$.
It is defined as follows: $\Gamma \mEnt A$ iff for every 
valuation $\nu$, either $\val{A'}{\nu} = \false$ for some 
$A' \in \Gamma$ or $\val{A}{\nu} \in \{\true,\both\}$.
We have that the Hilbert-style formulation of LP$^{\IImpl,\False}$ is 
strongly complete with respect to its semantics, i.e.\ $\Gamma \pEnt A$ 
iff $\Gamma \mEnt A$ (see e.g.~\cite{BC04a}).

The following properties of LP$^{\IImpl}$, shown in~\cite{AAZ11b},
suggest that LP$^{\IImpl}$ retains as much of classical propositional
logic as possible:
\begin{itemize}
\item
\emph{containment in classical logic}:
${\pEnt} \subseteq {\clpEnt}$;%
\footnote
{We write $\clpEnt$ for the logical consequence relation of classical 
 propositional logic.}
\item
\emph{proper implication}:
for all sets $\Gamma$ of formulas of LP$^{\IImpl}$ and all formulas $A$ 
and $B$ of LP$^{\IImpl}$: 
$\Gamma \cup \{A\} \pEnt B$ only if $\Gamma \pEnt A \IImpl B$;
\item
\emph{weakly maximal paraconsistency relative to classical logic}:
for all formulas $A$ of LP$^{\IImpl}$ with $\clpEnt A$ and not 
$\pEnt A$, for the minimal consequence relation $\extpEnt$ such that
${\pEnt} \subseteq {\extpEnt}$ and $\extpEnt A$, for all formulas $B$ 
of LP$^{\IImpl}$, $\extpEnt B$ iff $\clpEnt B$;%
\footnote
{Let $\vdash$ be a logical consequence relation.
 Then we write $\vdash A$ for $\emptyset \vdash A$.}
\item
\emph{strongly maximal absolute paraconsistency}:
for all propositional logics $\mathcal{L}$ with a consequence relation 
$\extpEnt$ such that ${\pEnt} \subset {\extpEnt}$, $\mathcal{L}$ is not 
paraconsistent.
\end{itemize}
These properties make LP$^{\IImpl}$ an ideal paraconsistent logic in the
sense made precise in~\cite{AAZ11b}.

We get the following logics if we enrich LP$^{\IImpl}$ with a constant
for $\false$, with a constant for $\both$ or with constants for both:
\begin{itemize}
\item
LP$^{\IImpl,\False}$ is LP$^{\IImpl}$ extended with a constant $\False$
and the axiom schema $\False \IImpl A$.
Valuations $\nu$ for LP$^{\IImpl,\False}$ are such that
$\val{\False}{\nu} = \false$.
\item
LP$^{\IImpl,\Both}$ is LP$^{\IImpl}$ extended with a constant $\Both$
and the axiom schemas $A \IImpl \Both$\linebreak[2] and
$A \IImpl \Not \Both$.
Valuations $\nu$ for LP$^{\IImpl,\Both}$ are such that
$\val{\Both}{\nu} = \both$.
\item
LP$^{\IImpl,\False,\Both}$ is LP$^{\IImpl}$ extended with constants
$\False$ and $\Both$ and the axiom schemas $\False \IImpl A$,
$A \IImpl \Both$ and $A \IImpl \Not \Both$.
Valuations $\nu$ for LP$^{\IImpl,\False,\Both}$ are such that
$\val{\False}{\nu} = \false$ and $\val{\Both}{\nu} = \both$.
\end{itemize}
The enrichments in question result in increase of expressive power.
Below, the properties of $\{\true,\false\}$-closure and
$\{\both\}$-freeness are used to characterize the expressive power of
the different logics:
\begin{itemize}
\item
a function $g$ from $\{\true,\false,\both\}^n$ to
$\{\true,\false,\both\}$ is $\{\true,\false\}$-\emph{closed} if the
image of the restriction of $g$ to $\{\true,\false\}^n$ is $\{\true\}$,
$\{\false\}$ or $\{\true,\false\}$;
\item
a function $g$ from $\{\true,\false,\both\}^n$ to
$\{\true,\false,\both\}$ is $\{\both\}$-\emph{free} if the image of the
re\-striction\nolinebreak[2] of $g$ to $\{\both\}^n$ is $\{\both\}$.
\end{itemize}
The expressive power of LP$^{\IImpl}$, LP$^{\IImpl,\False}$,
LP$^{\IImpl,\Both}$ and LP$^{\IImpl,\False,\Both}$ can now be
characterized as follows:
\begin{itemize}
\item
a function $g$ from $\{\true,\false,\both\}^n$ to
$\{\true,\false,\both\}$ is representable in the language
of\linebreak[2]
LP$^{\IImpl}$ iff it is $\{\true,\false\}$-closed and $\{\both\}$-free;
\item
a function $g$ from $\{\true,\false,\both\}^n$ to
$\{\true,\false,\both\}$ is representable in the language of
LP$^{\IImpl,\False}$ iff it is $\{\true,\false\}$-closed;
\item
a function $g$ from $\{\true,\false,\both\}^n$ to
$\{\true,\false,\both\}$ is representable in the language of
LP$^{\IImpl,\Both}$ iff it is $\{\both\}$-free;
\item
every function $g$ from $\{\true,\false,\both\}^n$ to
$\{\true,\false,\both\}$ is representable in the language of
LP$^{\IImpl,\False,\Both}$, i.e.\ the language of
LP$^{\IImpl,\False,\Both}$ is functionally complete.
\end{itemize}
With the exception of $\Or$ and $\And$, each of the connectives in
$\{\Not,\Or,\And,\IImpl,\False,\Both\}$ is not definable in terms of the
rest.
The connectives $\Or$ and $\And$ are definable in terms of
$\{\Not,\IImpl,\False,\Both\}$.
The preceding discussion of the expressive power of LP$^{\IImpl}$ and
some enrichments thereof is based on~\cite{Avr99a}.

Note that, in LP$^{\IImpl,\False}$ and LP$^{\IImpl,\False,\Both}$, a
constant $\True$ for $\true$ can simply be defined by
$\True = \Not \False$.
Note further that the consistency of a formula $A$ cannot be represented in
LP$^{\IImpl}$ and LP$^{\IImpl,\Both}$, but that it can be represented in
LP$^{\IImpl,\False}$ and LP$^{\IImpl,\False,\Both}$ by the formula
$(A \IImpl \False) \Or (\Not A \IImpl \False)$.%
\footnote
{In the setting of da Costa's logics C$_n$, which will be discussed in
 Section~\ref{sect-Cn}, the consistency of a formula is called the
 well-behavedness of a formula.
}
The properties that make the logic LP$^{\IImpl}$ an ideal paraconsistent 
logic in the sense of~\cite{AAZ11b} carry over to LP$^{\IImpl,\False}$,
LP$^{\IImpl,\Both}$ and LP$^{\IImpl,\False,\Both}$
(see e.g.~\cite{AAZ11b}).
LP$^{\IImpl,\False}$ is essentially the same logic as 
J$_3$~\cite{Dd70a,DOt85a} (see e.g.~\cite{CCM07a}).

\section{Interlude: LP$^{\IImpl}$ and its Dual}
\label{section-interlude}

It was mentioned in Section~\ref{sect-LPiimpl} that, if we add the axiom
schema $\Not A \IImpl (A \IImpl B)$ to the given Hilbert-style
formulation of LP$^{\IImpl}$, then we get a Hilbert-style for\-mulation
of classical propositional logic.
If we replace the axiom schema $A \Or \Not A$ by the axiom schema
$\Not A \IImpl (A \IImpl B)$ in the given Hilbert-style formulation of
LP$^{\IImpl}$ instead, then we get a Hilbert-style formulation of
Kleene's strong three-valued logic~\cite{Kle52a} enriched with an
implication connective for which the standard deduction theorem holds.
We use K$_3^{\IImpl}$ to denote this logic.
K$_3^{\IImpl}$ can be considered to be the dual of LP$^{\IImpl}$.
All differences between these two logics can be traced to the fact that
the third truth value $\both$ is interpreted as both true and false in
LP$^{\IImpl}$ and as neither true nor false in K$_3^{\IImpl}$.

Like in the case of LP$^{\IImpl}$, meanings are assigned to the
formulas of K$_3^{\IImpl}$ by means of valuations that are functions
from the set of all formulas of K$_3^{\IImpl}$ to the set
$\{\true,\false,\both\}$.
The conditions that a valuation for K$_3^{\IImpl}$ must satisfy differ
from the conditions that a valuation for LP$^{\IImpl}$ must satisfy only
with respect to implication:
\pagebreak[2]
\begin{eqnarray*}
\val{A \IImpl B}{\nu} & = &
 \left \{
 \begin{array}{l@{\;\;}l}
 \val{B}{\nu} & \mathrm{if}\; \val{A}{\nu} = \true \\
 \true        & \mathrm{otherwise}.
 \end{array}
 \right.
\end{eqnarray*}
The logical consequence relation of K$_3^{\IImpl}$, denoted by
$\Ent{\mathrm{K}_3^{\IImpl}}$, is the binary relation between sets of
formulas of K$_3^{\IImpl}$ and formulas of K$_3^{\IImpl}$ defined as
usual:
$ \Gamma \Ent{\mathrm{K}_3^{\IImpl}} A$ iff there exists a proof of $A$
from $\Gamma$ in K$_3^{\IImpl}$.
We have that $\Gamma \Ent{\mathrm{K}_3^{\IImpl}} A$ iff for every
valuation $\nu$, either $\val{A'}{\nu} \in \{\false,\both\}$ for some
$A' \in \Gamma$ or $\val{A}{\nu} = \true$.

In~\cite{AA96a}, a logic with four truth values ($\true$, $\false$,
$\both$ and $\neither$), called BL$_{\IImpl}$, is proposed in which both
LP$^{\IImpl}$ and K$_3^{\IImpl}$ can be simulated
(see e.g.~\cite{AA98a}).
This means in the case of LP$^{\IImpl}$ that
$\Gamma \Ent{\mathrm{LP}^{\IImpl}} A$ iff
$\Gamma \cup \{P_1 \Or \Not P_1,\ldots,P_n \Or \Not P_n\}
                                          \Ent{\mathrm{BL}_{\IImpl}} A$,
where $\{P_1,\ldots,P_n\}$ is the set of all propositional variables in
$\Gamma \cup \{A\}$.

\section{The Relevance Logic RM$_3$}
\label{sect-RM3}

The three-valued relevance-mingle logic RM$_3$~\cite{AB75a} is the
strongest logic among the logics that have been proposed by the school
of relevance logic.
Relevance logics are based on the idea that the antecedent of an
implication must be relevant to its consequent.
Although the meaning of relevance is nowhere made precise, it is a
characteristic feature of a relevance logic that propositions of the
forms $A \IImpl (B \IImpl A)$ and $A \IImpl (\Not A \IImpl B)$ do not
belong to its theorems.
This means that every relevance logic is a paraconsistent logic.

A Hilbert-style formulation of RM$_3$ is given in
Table~\ref{proofsystem-RM3}.
\begin{table}[!ht]
\caption{Hilbert-style formulation of RM$_3$}
\label{proofsystem-RM3}
\begin{eqntbl}
\begin{axcol}
\mathbf{Axiom\; Schemas:}
\\
A \IImpl A
\\
(A \IImpl B) \IImpl ((B \IImpl C) \IImpl (A \IImpl C))
\\
A \IImpl ((A \IImpl B) \IImpl B)
\\
(A \IImpl (A \IImpl B)) \IImpl (A \IImpl B)
\\
(A \And B) \IImpl A
\\
(A \And B) \IImpl B
\\
((A \IImpl B) \And (A \IImpl C)) \IImpl (A \IImpl (B \And C))
\\
A \IImpl (A \Or B)
\eqnsep
\mathbf{Rules\; of\; Inference:}
\\
\Infrule{A \quad A \IImpl B}{B}
\end{axcol}
\qquad
\begin{axcol}
{} \\
B \IImpl (A \Or B)
\\
((A \IImpl C) \And (B \IImpl C)) \IImpl ((A \Or B) \IImpl C)
\\
(A \And (B \Or C)) \IImpl ((A \And B) \Or (A \And C))
\\
\Not \Not A \IImpl A
\\
(A \IImpl \Not B) \IImpl (B \IImpl \Not A)
\\
{}
\\
A \IImpl (A \IImpl A)
\\
A \Or (A \IImpl B)
\eqnsep
{}
\\
\Infrule{A \quad B}{A \And B}
\end{axcol}
\end{eqntbl}
\end{table}
If we remove the last two axiom schemas on the right-hand side of
Table~\ref{proofsystem-RM3} from the given Hilbert-style formulation of
RM$_3$, then we get a Hilbert-style formulation of the well-known logic
R~\cite{AB75a}.
The axiom schema $A \IImpl (A \IImpl A)$ is known as the mingle axiom.
If we add the axiom schema $A \IImpl (B \IImpl A)$, which generalizes
the mingle axiom, to the given Hilbert-style formulation of RM$_3$, then
we get a Hilbert-style formulation of classical propositional logic
(see e.g.~\cite{Dun00a}).

The following outline of the semantics of RM$_3$ is based
on~\cite{Avr91a}.
Like in the case of LP$^{\IImpl}$, meanings are assigned to the formulas
of RM$_3$ by means of valuations that are functions from the set of all
formulas of RM$_3$ to the set $\{\true,\false,\both\}$.
The conditions that a valuation for RM$_3$ must satisfy differ from the
conditions that a valuation for LP$^{\IImpl}$ must satisfy only with
respect to implication:
\begin{eqnarray*}
\val{A \IImpl B}{\nu} & = &
 \left \{
 \begin{array}{l@{\;\;}l}
 \both  & \mathrm{if}\;
          \val{A}{\nu} = \both  \;\mathrm{and}\; \val{B}{\nu} = \both \\
 \true  & \mathrm{if}\;
          \val{A}{\nu} = \false \;\mathrm{or}\;  \val{B}{\nu} = \true \\
 \false & \mathrm{otherwise}.
 \end{array}
 \right.
\end{eqnarray*}

The logical consequence relation of RM$_3$ is also based on the idea
that a valuation $\nu$ satisfies a formula $A$ if
$\val{A}{\nu} \in \{\true,\both\}$, but is adapted to the different kind
of implication found in RM$_3$.
We have that $\Gamma \Ent{\mathrm{RM}_3} A$ iff
for every valuation $\nu$,
either $\val{A'}{\nu} = \false$ for some $A' \in \Gamma$,
or $\val{A}{\nu} = \true$,
or $\val{A'}{\nu} = \both$ for all $A' \in \Gamma$
and $\val{A}{\nu} = \both$.

By results from~\cite{AAZ11b,AAZ11a}, it is easy to see that RM$_3$ has
the four properties that make it an ideal paraconsistent logic in the
sense of~\cite{AAZ11b} as well.

The implication connective of LP$^{\IImpl}$, here written $\IImpl_*$, can
be defined in RM$_3$~by
\begin{eqnarray*}
A \IImpl_* B = B \Or (A \IImpl B)
\end{eqnarray*}
and the other way round, the implication connective $\IImpl$ of RM$_3$
can be defined in LP$^{\IImpl}$ by
\begin{eqnarray*}
A \IImpl B = (A \IImpl_* B) \And (\Not B \IImpl_* \Not A)
\end{eqnarray*}
(see e.g.~\cite{Avr91a}).
Hence, RM$_3$ has the same expressive power as LP$^{\IImpl}$.
To increase the expressive power, RM$_3$ can be enriched with a constant
for $\false$ in the same way as LP$^{\IImpl}$ is enriched with a
constant for $\false$ in Section~\ref{sect-LPiimpl}.
In the resulting logic, like in LP$^{\IImpl,\False}$, the consistency of
a formula $A$ can be represented by the formula
$(A \IImpl \False) \Or (\Not A \IImpl \False)$.

\section{Da Costa's Paraconsistent Logics C$_n$}
\label{sect-Cn}

In~\cite{daC74a}, da Costa proposes the paraconsistent propositional
logics C$_n$, for $n > 0$.
In these logics, paraconsistency is obtained by adopting weak forms of
negations.
These forms of negation are, to a certain extent, duals of the weak form
of negation found in intuitionistic logic: if something is false then
its negation must be true, but if something is true then its negation
may be true as well.
In the case that something is true, further conditions are imposed, but
they never have the effect that its negation must be true.
In this way, the logics in question allow for contradictory formulas to
be true.

A formula of the form $A \And \Not A$ is a contradictory formula.
If the formula $\Not (A \And \Not A)$ is true, then $A$ is called a
well-behaved formula of degree $1$; if in addition the formula
$\Not (\Not (A \And \Not A) \And \Not \Not (A \And \Not A))$ is
true, then $A$ is called a well-behaved formula of degree $2$; etc.
We introduce abbreviations to express that a formula is a well-behaved
formula of degree $n$, for $n > 0$.
The abbreviations $A\Con{n}$, for $n > 0$, are recursively defined as
follows:
$A\Con{1}$ stands for $A^1$,
$A\Con{n+1}$ stands for $A\Con{n} \And A^{n+1}$, where
the auxiliary abbreviations $A^n$, for $n \geq 0$, are recursively
defined as follows:
$A^0$ stands for $A$,
$A^{n+1}$ stands for $\Not (A^n \And \Not A^n)$.

A Hilbert-style formulation of C$_n$ is given in
Table~\ref{proofsystem-Cn}.
\begin{table}[!ht]
\caption{Hilbert-style formulation of C$_n$}
\label{proofsystem-Cn}
\begin{eqntbl}
\begin{axcol}
\mathbf{Axiom\; Schemas:}
\\
A \IImpl (B \IImpl A)
\\
(A \IImpl B) \IImpl ((A \IImpl (B \IImpl C)) \IImpl (A \IImpl C))
\\
(A \And B) \IImpl A
\\
(A \And B) \IImpl B
\\
A \IImpl (B \IImpl (A \And B))
\\
A \IImpl (A \Or B)
\\
B \IImpl (A \Or B)
\\
(A \IImpl C) \IImpl ((B \IImpl C) \IImpl ((A \Or B) \IImpl C))
\eqnsep
\mathbf{Rule\; of\; Inference:}
\\
\Infrule{A \quad A \IImpl B}{B}
\end{axcol}
\qquad
\begin{axcol}
{} \\
A \Or \Not A
\\
\Not \Not A \IImpl A
\\
B\Con{n} \IImpl ((A \IImpl B) \IImpl ((A \IImpl \Not B) \IImpl \Not A))
\\
(A\Con{n} \And B\Con{n}) \IImpl (A \IImpl B)\Con{n}
\\
(A\Con{n} \And B\Con{n}) \IImpl (A \And B)\Con{n}
\\
(A\Con{n} \And B\Con{n}) \IImpl (A \Or B)\Con{n}
\end{axcol}
\end{eqntbl}
\end{table}
The axiom schemas on the left-hand side of
Table~\ref{proofsystem-Cn} and the single inference rule (modus
ponens) constitute a Hilbert-style formulation of the positive fragment
of intuitionistic propositional logic.
The third axiom schema on the right-hand side expresses that the version
of reductio ad absurdum that is found in a Hilbert-style formulation of
full of intuitionistic propositional logic, viz.\
$(A \IImpl B) \IImpl ((A \IImpl \Not B) \IImpl \Not A)$,
works provided that $B$ is a well-behaved formula of degree $n$.
The last three axiom schemas say that formulas composed of well-behaved
formulas of degree $n$ are well-behaved formulas of degree $n$ as well.
If we add the axiom schema $\Not (A \And \Not A)$, which is called the
law of noncontradiction, to the given Hilbert-style formulation of
C$_n$, then we get a Hilbert-style formulation of classical
propositional logic (see e.g.~\cite{dKB07a}).

The following outline of the semantics of C$_n$ is based
on~\cite{daC77a}.
Like in the cases of LP$^{\IImpl}$ and RM$_3$, meanings are assigned to
the formulas of C$_n$ by means of valuations.
However, different from valuations for LP$^{\IImpl}$ and RM$_3$,
valuations for C$_n$ are functions from the set of all formulas of C$_n$
to the set $\{\true,\false\}$ (like in the case of classical
propositional logic).
The conditions that a valuation for C$_n$ must satisfy differ from the
conditions that a valuation for classical propositional logic must
satisfy only with respect to negation:
\pagebreak[2]
\begin{ldispl}
\begin{aeqns}
\val{\Not A}{\nu} = \true              & \mathrm{if} &
 \val{A}{\nu} = \false \\
\val{A}{\nu} = \true                   & \mathrm{if} &
 \val{\Not \Not A}{\nu} = \true \\
\val{A}{\nu} = \false                  & \mathrm{if} &
 \val{A\Con{n}}{\nu} = \true \;\mathrm{and}\;
 \val{A \IImpl B}{\nu} = \true \;\mathrm{and}\;
 \val{A \IImpl \Not B}{\nu} = \true \\
\val{(A \IImpl B)\Con{n}}{\nu} = \true & \mathrm{if} &
 \val{A\Con{n}}{\nu} = \true \;\mathrm{and}\;
 \val{B\Con{n}}{\nu} = \true \\
\val{(A \And B)\Con{n}}{\nu}   = \true & \mathrm{if} &
 \val{A\Con{n}}{\nu} = \true \;\mathrm{and}\;
 \val{B\Con{n}}{\nu} = \true \\
\val{(A \Or B)\Con{n}}{\nu}    = \true & \mathrm{if} &
 \val{A\Con{n}}{\nu} = \true \;\mathrm{and}\;
 \val{B\Con{n}}{\nu} = \true.
\end{aeqns}
\end{ldispl}
With these unusual conditions, there exist valuations that assign the
truth value~$\true$ to at least one contradictory formula.
Clearly, these conditions are nothing else but semantic counterparts of
the axiom schemas in which the negation connective occurs.
Therefore, they do not help in gaining a better insight into the
negation connective of C$_n$.
At best, these conditions confirm the feeling that the negation
connective of C$_n$ is not really a contrary forming operator.
By the unusual conditions, unlike the valuations for LP$^{\IImpl}$ and
RM$_3$, the valuations for C$_n$ are not fully determined by the truth
values that they assign to the propositional variables.
That is, negation is made non-truth-functional in C$_n$.

Like the logical consequence relation of classical propositional logic,
the logical consequence relation of C$_n$ is based on the idea that a
valuation $\nu$ satisfies a formula $A$ if $\val{A}{\nu} = \true$.
We have that $\Gamma \Ent{\mathrm{C}_n} A$ iff
for every valuation $\nu$,
either $\val{A'}{\nu} = \false$ for some $A' \in \Gamma$
or $\val{A}{\nu} = \true$.

The logical consequence relation of C$_n$ is included in the logical
consequence relation of classical propositional logic and the
implication connective of C$_n$ is such that the classical deduction
theorem holds (see e.g.~\cite{dKB07a}).
However, maximal paraconsistency relative to classical logic and
absolute strongly maximal paraconsistency are not properties of C$_n$
(see e.g.~\cite{CCM07a,Avr05a}).
Consequently, C$_n$ is not an ideal paraconsistent logic in the sense
of~\cite{AAZ11b}.

Although the negation connective of C$_n$ is non-truth-functional, the
truth-functional negation connective of classical propositional logic,
here written $\Not_*$, can be defined in C$_n$ by
$\Not_* A = \Not A \And A\Con{n}$ (see e.g.~\cite{daC77a}).

Little is known about the connections between the logics C$_n$ and the
other paraconsistent logics discussed in this note.
In~\cite{CCM07a}, Carnielli and others introduce several classes of
paraconsistent logics, including the class of LFIs (Logics of Formal
Inconsistency) and the class of dC-systems.
The class of dC-systems is a subclass of the class of LFIs.
The logics C$_n$ are dC-systems.
The logic LP$^{\IImpl}$ is not even an LFI, because connectives for
consistency and inconsistency must be definable in an LFI.
However, the logics LP$^{\IImpl,\False}$ and LP$^{\IImpl,\False,\Both}$
are dC-systems.
An interesting collection of thousands of dC-systems with maximal
paraconsistency relative to classical logic and absolute strongly
maximal paraconsistency are identified in~\cite{CCM07a}.
Any of the dC-systems in question can be conservatively translated into
LP$^{\IImpl,\False}$ and LP$^{\IImpl,\False,\Both}$.%
\footnote
{For a characterization of the dC-systems in question and a definition
 of conservative translation, see~\cite{CCM07a}.
}
However, because maximal paraconsistency relative to classical logic and
absolute strongly maximal paraconsistency are not properties of them,
the logics C$_n$ do not belong to the collection.

\section{Jaskow\'{s}ki's Paraconsistent Logic D$_2$}
\label{sect-D2}

In~\cite{Jas99a,Jas99b}, Ja\'{s}kowski proposes the paraconsistent logic
D$_2$.%
\footnote
{The cited publications are translations of Polish editions published in
 1948 and 1949.
}
Ja\'{s}kowski calls this logic a discussive logic.
His basic idea is to take true as true according to the position
of some person engaged in a discussion.
True in this sense can be thought of as true in some possible world,
namely the world of some person's position.
Thus, both $A$ and $\Not A$ can be true without an arbitrary formula $B$
being true.
Initially, D$_2$ was presented as a modal logic in disguise.

A Hilbert-style formulation of D$_2$ is given in
Table~\ref{proofsystem-D2}.
\begin{table}[!ht]
\caption{Hilbert-style formulation of D$_2$}
\label{proofsystem-D2}
\begin{eqntbl}
\begin{axcol}
\mathbf{Axiom\; Schemas:}
\\
A \IImpl (B \IImpl A)
\\
(A \IImpl (B \IImpl C)) \IImpl ((A \IImpl B) \IImpl (A \IImpl C))
\\
(A \And B) \IImpl A
\\
(A \And B) \IImpl B
\\
(A \IImpl B) \IImpl ((A \IImpl C) \IImpl (A \IImpl (B \And C)))
\\
A \IImpl (A \Or B)
\\
B \IImpl (A \Or B)
\\
(A \IImpl C) \IImpl ((B \IImpl C) \IImpl ((A \Or B) \IImpl C))
\\
A \Or (A \IImpl B)
\eqnsep
\mathbf{Rule\; of\; Inference:}
\\
\Infrule{A \quad A \IImpl B}{B}
\end{axcol}
\qquad
\begin{axcol}
{}
\\
\Not (\Not A \And \Not \Not A \And \Not (A \Or \Not A))
\\
\Not (\Not A \And \Not B \And \Not (A \Or B))
 \IImpl {} \\ \qquad
\Not (\Not A \And \Not B \And \Not C \And \Not (A \Or B \Or C))
\\
\Not (\Not A \And \Not B \And \Not C \And \Not (A \Or B \Or C))
 \IImpl {} \\ \qquad
\Not (\Not A \And \Not C \And \Not B \And \Not (A \Or C \Or B))
\\
\Not (\Not A \And \Not B \And \Not C \And \Not (A \Or B \Or C))
 \IImpl {} \\ \qquad
((A \Or B \Or \Not C) \IImpl (A \Or B))
\\
\Not (\Not A \And \Not B) \IImpl (A \Or B)
\\
(A \Or (B \Or \Not B)) \IImpl \Not (\Not A \And \Not (B \Or \Not B))
\end{axcol}
\end{eqntbl}
\end{table}
This formulation is taken from~\cite{Ciu08a}.
Various axiom schemas from Table~\ref{proofsystem-D2} are rather unusual
as axiom schemas.
However, they are all schemas of tautologies of classical propositional
logic.

The following outline of the semantics of D$_2$ is based
on~\cite{Ciu08a}.
Unlike in the cases of LP$^{\IImpl}$, RM$_3$ and C$_n$, meanings are
assigned to the formulas of D$_2$ by means of pairs $(W,\nu)$ where $W$
is a non-empty set of \emph{worlds} and $\nu$ is a function from the
cartesian product of the set of all formulas of D$_2$ and the set $W$ to
the set $\{\true,\false\}$ such that for all formulas $A$ and $B$ of
D$_2$:
\begin{eqnarray*}
\val{A \IImpl B,w}{\nu} & = &
 \left \{
 \begin{array}{l@{\;\;}l}
 \true  & \mathrm{if}\; \mathrm{for\; all}\;  w' \in W,\;
                        \val{A,w'}{\nu} = \false \;\mathrm{or}\;
                        \val{B,w}{\nu}  = \true  \\
 \false & \mathrm{if}\; \mathrm{for\; some}\; w' \in W,\;
                        \val{A,w'}{\nu} = \true  \;\mathrm{and}\;
                        \val{B,w}{\nu}  = \false,
 \end{array}
 \right.
\\
\val{A \And B,w}{\nu} & = &
 \left \{
 \begin{array}{l@{\;\;}l}
 \true  & \mathrm{if}\; \val{A,w}{\nu}  = \true  \;\mathrm{and}\;
                        \mathrm{for\; some}\; w' \in W,\;
                        \val{B,w'}{\nu} = \true  \\
 \false & \mathrm{if}\; \val{A,w}{\nu}  = \false \;\mathrm{or}\;
                        \mathrm{for\; all}\;  w' \in W,\;
                        \val{B,w'}{\nu} = \false,
 \end{array}
 \right.
\\
\val{A \Or B,w}{\nu} & = &
 \left \{
 \begin{array}{l@{\;\;}l}
 \true  & \mathrm{if}\; \val{A,w}{\nu} = \true  \;\mathrm{or}\;
                        \val{B,w}{\nu} = \true  \\
 \false & \mathrm{if}\; \val{A,w}{\nu} = \false \;\mathrm{and}\;
                        \val{B,w}{\nu} = \false,
 \end{array}
 \right.
\\
\val{\Not A,w}{\nu} & = &
 \left \{
 \begin{array}{l@{\;\;}l}
 \true  & \mathrm{if}\; \val{A,w}{\nu} = \false \\
 \false & \mathrm{if}\; \val{A,w}{\nu} = \true.
 \end{array}
 \right.
\end{eqnarray*}
These pairs are called discussive structures.
A discussive structure is essentially the same as a Kripke structure of
which the accessibility relation includes every pair of worlds.
The truth-conditions and falsehood-conditions for the logical
connectives become the classical ones if the number of worlds is
restricted to one.
The conditions for the implication connective and the conjunction
connective reveal their modal nature clearly.

The logical consequence relation of D$_2$ is based on the idea that a
discussive structure $(W,\nu)$ satisfies a formula $A$ if
$\val{A,w}{\nu} = \true$ for some $w \in W$.
We have that $\Gamma \Ent{\mathrm{D}_2} A$ iff
for every discussive structure $(W,\nu)$,
either $\val{A',w'}{\nu} = \false$ for all $w' \in W$ for some
$A' \in \Gamma$
or $\val{A,w}{\nu} = \true$ for some $w \in W$.

The logical consequence relation of D$_2$ is included in the logical
consequence relation of classical propositional logic and the
implication connective of D$_2$ is such that the classical deduction
theorem holds (see e.g.~\cite{Ciu08a}).
It is unknown to me whether maximal paraconsistency relative to
classical logic and/or absolute strongly maximal paraconsistency are
properties of D$_2$ and consequently whether D$_2$ is an ideal
paraconsistent logic in the sense of~\cite{AAZ11b}.

Little is known about the connections between the logic D$_2$ and the
other paraconsistent logics discussed in this note.
Like LP$^{\IImpl}$, D$_2$ is not even an LFI.
If D$_2$ is enriched with the necessity operator $\Box$ satisfying the
axioms of S5, which is the modal operator used in~\cite{Jas99a,Jas99b}
to explain D$_2$, we get an LFI.

\section{The Annotated Logics P$\tau$}
\label{sect-Ptau}

In~\cite{Sub87a}, Subrahmanian takes the first step towards the
paraconsistent propositional logics called P$\tau$~\cite{dSV91a}.
Here, $\tau$ is some triple $(|\tau|,\leq,\mnot)$, where $(|\tau|,\leq)$
is a complete lattice of (object-level) truth values and $\mnot$ is a
function $\mnot \colon |\tau| \to |\tau|$ that gives the meaning of
negation in P$\tau$.
A typical case is the one where
$|\tau| = \{\mneither,\mtrue,\mfalse,\mboth\}$, $\mneither \leq x$,
$x \leq \mboth$, $x \not\leq \mnot(x)$ if $x \in \{\mtrue,\mfalse\}$,
$\mnot(\mtrue) = \mfalse$, $\mnot(\mfalse) = \mtrue$, and $\mnot(x) = x$
if $x \in \{\mneither,\mboth\}$.
P$\tau$ is called an annotated logic.
In P$\tau$, propositional variables are annotated with an element from
$|\tau|$.
An annotated propositional variable $P_\lambda$, where $P$ is an
ordinary propositional variable and $\lambda \in |\tau|$, expresses that
it is believed that $P$'s truth value is at least $\lambda$.

We use the symbols $\bot$ and $\top$ to denote the bottom element and
the top element, respectively, of the complete lattice $(|\tau|,\leq)$.
Moreover, we write $\bigsqcup_{i=1}^n \lambda_i$, where
$\lambda_1,\ldots,\lambda_n \in |\tau|$, for the least upper bound of
the set $\{\lambda_1,\ldots,\lambda_n\}$ with respect to $\leq$.

We also introduce abbreviations for multiple negations.
The abbreviations $\Not^n A$, for $n \geq 0$, are recursively defined as
follows:
$\Not^0 A$ stands for $A$ and
$\Not^{n+1} A$ stands for $\Not (\Not^n A)$.
A formula which is not of the form $\Not^n P_\lambda$, where $P_\lambda$
is an annotated propositional variable, is called a complex formula.

A Hilbert-style formulation of P$\tau$ is given in
Table~\ref{proofsystem-Ptau}.
\begin{table}[!ht]
\caption{Hilbert-style formulation of P$\tau$}
\label{proofsystem-Ptau}
\begin{eqntbl}
\begin{axcol}
\mathbf{Axiom\; Schemas:}
\\
A \IImpl (B \IImpl A)
\\
(A \IImpl (B \IImpl C)) \IImpl ((A \IImpl B) \IImpl (A \IImpl C))
\\
((A \IImpl B) \IImpl A) \IImpl A
\\
(A \And B) \IImpl A
\\
(A \And B) \IImpl B
\\
A \IImpl (B \IImpl (A \And B))
\\
A \IImpl (A \Or B)
\\
B \IImpl (A \Or B)
\\
(A \IImpl C) \IImpl ((B \IImpl C) \IImpl ((A \Or B) \IImpl C))
\eqnsep
\mathbf{Rule\; of\; Inference:}
\\
\Infrule{A \quad A \IImpl B}{B}
\end{axcol}
\qquad
\begin{axcol}
{}
\\
(F \IImpl G) \IImpl ((F \IImpl \Not G) \IImpl \Not F)
\\
F \IImpl (\Not F \IImpl A)
\\
F \Or \Not F
\\
{}
\\
P_\mbot
\\
\Not^{n+1} P_\lambda \IIff \Not^n P_{\mnot(\lambda)}
\\
P_\lambda \IImpl P_\mu \quad\mathrm{if}\; \lambda \geq \mu
\\
P_{\lambda_1} \And \ldots \And P_{\lambda_n} \IImpl P_\lambda
 \quad\mathrm{if}\; \lambda = \bigsqcup_{i=1}^n \lambda_i
\end{axcol}
\end{eqntbl}
\end{table}
In this table,
$F$ and $G$ range over all complex formulas of P$\tau$,
$P$ ranges over propositional variables, and
$\lambda,\mu,\lambda_1,\ldots,\lambda_n$ range over $|\tau|$.
This formulation is based on the formulation of Q$\tau$, the first-order
counterpart of P$\tau$, given in~\cite{ANA07a}.
The axiom schemas on the left-hand side of Table~\ref{proofsystem-Ptau}
and the single inference rule (modus ponens) constitute a Hilbert-style
formulation of the positive fragment of classical propositional logic.
The first three axiom schemas on the right-hand side of
Table~\ref{proofsystem-Ptau} are the usual axiom schemas for negation in
a Hilbert-style formulation of classical propositional logic, but here
their possible instances are restricted.
The last four axiom schemas on the right-hand side of
Table~\ref{proofsystem-Ptau} are special axiom schemas concerning
formulas of the forms $P_\lambda$ and $\Not^{n+1} P_\lambda$.

The following outline of the semantics of P$\tau$ is based
on~\cite{ANA07a}.
Like in the cases of LP$^{\IImpl}$, RM$_3$ and C$_n$, meanings are
assigned to the formulas of P$\tau$ by means of valuations.
Like in the cases of C$_n$, valuations for P$\tau$ are functions from
the set of all formulas of P$\tau$ to the set $\{\true,\false\}$.
A valuation for P$\tau$ is such that for all formulas $A$ and $B$ of
P$\tau$, all complex formulas $F$ of P$\tau$, and all propositional
variables $P$:
\begin{eqnarray*}
\val{A \IImpl B}{\nu} & = &
 \left \{
 \begin{array}{l@{\;\;}l}
 \true  & \mathrm{if}\; \val{A}{\nu} = \false  \;\mathrm{or}\;
                        \val{B}{\nu} = \true  \\
 \false & \mathrm{if}\; \val{A}{\nu} = \true \;\mathrm{and}\;
                        \val{B}{\nu} = \false,
 \end{array}
 \right.
\\
\val{A \And B}{\nu} & = &
 \left \{
 \begin{array}{l@{\;\;}l}
 \true  & \mathrm{if}\; \val{A}{\nu} = \true  \;\mathrm{and}\;
                        \val{B}{\nu} = \true  \\
 \false & \mathrm{if}\; \val{A}{\nu} = \false \;\mathrm{or}\;
                        \val{B}{\nu} = \false,
 \end{array}
 \right.
\\
\val{A \Or B}{\nu} & = &
 \left \{
 \begin{array}{l@{\;\;}l}
 \true  & \mathrm{if}\; \val{A}{\nu} = \true  \;\mathrm{or}\;
                        \val{B}{\nu} = \true  \\
 \false & \mathrm{if}\; \val{A}{\nu} = \false \;\mathrm{and}\;
                        \val{B}{\nu} = \false,
 \end{array}
 \right.
\\
\val{\Not F}{\nu} & = &
 \left \{
 \begin{array}{l@{\;\;}l}
 \true  & \mathrm{if}\; \val{F}{\nu} = \false \\
 \false & \mathrm{if}\; \val{F}{\nu} = \true,
 \end{array}
 \right.
\\
\val{\Not^{n+1} P_\lambda}{\nu} & = &
 \left \{
 \begin{array}{l@{\;\;}l}
 \true  & \mathrm{if}\; \val{\Not^n P_{\mnot(\lambda)}}{\nu} = \true \\
 \false & \mathrm{if}\; \val{\Not^n P_{\mnot(\lambda)}}{\nu} = \false,
 \end{array}
 \right.
\\
\val{P_\lambda}{\nu} & = &
 \left \{
 \begin{array}{l@{\;\;}l}
 \true  & \mathrm{if}\; \mathrm{for\; some}\; \mu \geq \lambda,\;
          \val{P_\mu}{\nu} = \true \\
 \false & \mathrm{if}\; \mathrm{for\; all}\;  \mu \geq \lambda,\;
          \val{P_\mu}{\nu} = \false.
 \end{array}
 \right.
\end{eqnarray*}
The classical truth-conditions and falsehood-conditions for the logical
connectives are retained except for the negation connective.
There are special conditions for the occurrences of the negation
connective in formulas of the form $\Not^{n+1} P_\lambda$.

Like the logical consequence relation of C$_n$, the logical consequence
relation of P$\tau$ is based on the idea that a valuation $\nu$ satisfies
a formula $A$ if $\val{A}{\nu} = \true$.
We have that $\Gamma \Ent{\mathrm{P}\tau} A$ iff
for every valuation $\nu$,
either $\val{A'}{\nu} = \false$ for some $A' \in \Gamma$
or $\val{A}{\nu} = \true$.

The properties that make a paraconsistent logic ideal in the sense
of~\cite{AAZ11b} are unmeaning in the case of P$\tau$, because its
language is substantially deviating due to the annotated propositional
variables.

The negation connective of classical propositional logic, here written
$\Not_*$, can be defined in P$\tau$ by
$\Not_* A = A \IImpl ((A \IImpl A) \And \Not (A \IImpl A))$
(see e.g.~\cite{ANA07a}).

It seems that nothing is known about the connections between the logic
P$\tau$ and the other paraconsistent logics discussed in this note.
This is not completely unexpected, because the language of P$\tau$ is
substantially deviating.

\section{Paraconsistent Deontic Logics}
\label{sect-deontic-logics}

Legal reasoning is reasoning in the presence of inconsistent information
and reasoning concerned with normative expressions such as obligatory,
permissible, and prohibited.
This calls for a paraconsistent deontic logic.
Because paraconsistent deontic logics have not been investigated
extensively, only one is discussed in some detail here.
The logic concerned is called DLP$^{\IImpl,\False}$ because it is a
deontic logic based on LP$^{\IImpl,\False}$.
It is essentially the same logic as the deontic logic DLFI1 proposed
in~\cite{Con07a}.

DLP$^{\IImpl,\False}$ has the connectives of LP$^{\IImpl,\False}$ and in
addition an obligation connective $\Oblig$.
A formulation of this logic is obtained by adding to the Hilbert-style
formulation of LP$^{\IImpl}$ given in Table~\ref{proofsystem-LPiimpl}
the axiom schema $\False \IImpl A$ and the deontic axiom schemas and
rule of inference given in Table~\ref{add-DLPiimpl}.
\begin{table}[!ht]
\caption{Deontic axiom schemas and rule of inference for
  DLP$^{\IImpl,\False}$}
\label{add-DLPiimpl}
\begin{eqntbl}
\begin{axcol}
\mathbf{Axiom\; Schemas:}
\\
\Oblig (A \IImpl B) \IImpl (\Oblig A \IImpl \Oblig B)
\\
\Oblig \False \IImpl \False
\eqnsep
\mathbf{Rule\; of\; Inference:}
\\
\Infrule{A \;\mathrm{is\;a\;theorem}}
        {\Oblig A \;\mathrm{is\;a\;theorem}}
\end{axcol}
\end{eqntbl}
\end{table}
Of course, the additional rule of inference does not belong in a genuine
Hilbert-style formulation of a logic.
Like in LP$^{\IImpl}$, the standard deduction theorem holds in
DLP$^{\IImpl,\False}$.

Unlike in the case of LP$^{\IImpl}$, but like in the case of other
modal logics, meanings are assigned to the formulas of
DLP$^{\IImpl,\False}$ by means of triples $(W,R,\nu)$ where
$W$ is a non-empty set of \emph{possible worlds},
$R \subseteq W \times W$ is an \emph{accessibility} relation for which
it holds that for all $w \in W$ there exists a $w' \in W$ such that
$w R w'$, and
$\nu$ is a function from the cartesian product of the set of all
formulas of DLP$^{\IImpl,\False}$ and the set $W$ to the set
$\{\true,\false,\both\}$ such that for all formulas $A$ and $B$ of
DLP$^{\IImpl,\False}$:
\pagebreak[2]
\begin{eqnarray*}
\val{A \IImpl B,w}{\nu} & = &
 \left \{
 \begin{array}{l@{\;\;}l}
 \true        & \mathrm{if}\; \val{A,w}{\nu} = \false \\
 \val{B,w}{\nu} & \mathrm{otherwise},
 \end{array}
 \right.
\\
\val{A \And B,w}{\nu} & = &
 \left \{
 \begin{array}{l@{\;\;}l}
 \true  & \mathrm{if}\; \val{A,w}{\nu} = \true  \;\mathrm{and}\;
                        \val{B,w}{\nu} = \true  \\
 \false & \mathrm{if}\; \val{A,w}{\nu} = \false \;\mathrm{or}\;
                        \val{B,w}{\nu} = \false \\
 \both  & \mathrm{otherwise},
 \end{array}
 \right.
\\
\val{A \Or B,w}{\nu} & = &
 \left \{
 \begin{array}{l@{\;\;}l}
 \true  & \mathrm{if}\; \val{A,w}{\nu} = \true  \;\mathrm{or}\;
                        \val{B,w}{\nu} = \true  \\
 \false & \mathrm{if}\; \val{A,w}{\nu} = \false \;\mathrm{and}\;
                        \val{B,w}{\nu} = \false \\
 \both  & \mathrm{otherwise},
 \end{array}
 \right.
\\
\val{\Not A,w}{\nu} & = &
 \left \{
 \begin{array}{l@{\;\;}l}
 \true  & \mathrm{if}\; \val{A,w}{\nu} = \false \\
 \false & \mathrm{if}\; \val{A,w}{\nu} = \true \\
 \both  & \mathrm{otherwise},
 \end{array}
 \right.
\\
\val{\False,w}{\nu} & = & \false,
\\
\val{\Oblig A,w}{\nu} & = &
 \left \{
 \begin{array}{l@{\;\;}l}
 \true  & \mathrm{if}\; \mathrm{for\; all}\;  w' \in W \;\mathrm{with}\;
                        w R w',\; \val{A,w'}{\nu} = \true  \\
 \false & \mathrm{if}\; \mathrm{for\; some}\; w' \in W \;\mathrm{with}\;
                        w R w',\; \val{A,w'}{\nu} = \false \\
 \both  & \mathrm{otherwise}.
 \end{array}
 \right.
\end{eqnarray*}
These triples are called three-valued Kripke structures.
The truth-conditions and falsehood-conditions for the logical
connectives $\IImpl$, $\And$, $\Or$ and $\Not$ are the ones used in the
semantics of LP$^{\IImpl}$ for each possible world.
The conditions for the obligation connective reveals its modal nature
clearly.

The logical consequence relation of DLP$^{\IImpl,\False}$ is based on
the idea that a three-valued Kripke structure $(W,R,\nu)$ satisfies a
formula $A$ if $\val{A,w}{\nu} \in \{\true,\both\}$ for all $w \in W$.
We have that $\Gamma \Ent{\mathrm{DLP}^{\IImpl,\False}} A$ iff
for every three-valued Kripke structure $(W,R,\nu)$ and $w \in W$,
either $\val{A',w}{\nu} = \false$ for some $A' \in \Gamma$ or
$\val{A,w}{\nu} \in \{\true,\both\}$.

DLP$^{\IImpl,\False}$ is a \emph{deontically paraconsistent} logic,
i.e.\ there exist sets $\Gamma$ of formulas of DLP$^{\IImpl,\False}$ and
formulas $A$ of DLP$^{\IImpl,\False}$ such that not for all formulas $B$
of DLP$^{\IImpl,\False}$,
$\Gamma \cup \{\Oblig A,\Oblig \Not A\}
  \Ent{\mathrm{DLP}^{\IImpl,\False}} \Oblig B$.
A formula $A$ of DLP$^{\IImpl,\False}$ is called \emph{deontically
inconsistent} if
$\Oblig \Not ((A \IImpl \False) \Or (\Not A \IImpl \False))$.
From contradictory obligations $\Oblig A$ and $\Oblig \Not A$, it can be
inferred that $A$ is deontically inconsistent: for all sets $\Gamma$ of
formulas of DLP$^{\IImpl,\False}$ and all formulas $A$ of
DLP$^{\IImpl,\False}$, we have that
$\Gamma \Ent{\mathrm{DLP}^{\IImpl,\False}}
 \Oblig A \IImpl (\Oblig \Not A \IImpl
           \Oblig \Not ((A \IImpl \False) \Or (\Not A \IImpl \False)))$.

Deontic logics based on RM$_3$ and C$_1$ are devised in a similar way
in~\cite{McG06a} and~\cite{CC86a}, respectively.%
\footnote
{In fact, a family of modal logics based on RM$_3$, including a deontic
logic is devised in~\cite{McG06a}.
}
Published work on deontic logics based on D$_2$ and P$\tau$ seems to be
non-existent.
As far as modal logics based on P$\tau$ are concerned, all published
work seems to be on epistemic logics (see e.g.~\cite{AN09a}).
In fact, virtually all published work on paraconsistent modal logics
seems to be on epistemic logics based on P$\tau$.

\section{Belief Revision and Related Issues}
\label{sect-miscellaneous}

It is reasonable to assume that legal reasoning is done by rational
persons who revise their beliefs in the light of new information.
The AGM theory of belief revision~\cite{AGM85a,Gar88a} is a theory about
the dynamics of the beliefs of a rational person that is based on the
representation of beliefs as formulas of some logic.
Most of the work on belief revision is based on the AGM theory.
Much of this work takes for granted that the beliefs of a rational
person must be consistent.
However, motivated by the need to account for belief revision in the
presence of inconsistent beliefs, some work has been done on the
adaptation of the AGM theory to paraconsistent logics.

In~\cite{RS95a}, the AGM theory is adapted to a fragment of the logic
BL$_{\IImpl}$ mentioned at the end of Section~\ref{section-interlude}
enriched with constants for $\true$ and $\false$ and it is shown that
the adaptation has only minor consequences.
A similar adaptation is found in~\cite{LL96a}.
In~\cite{GRR08a}, it is shown how an adaptation of the AGM theory to the
same fragment of BL$_{\IImpl}$ can be obtained via a translation to
classical logic.
In~\cite{CB98a}, the AGM theory is adapted to the logics C$_n$.
In~\cite{Mar02a}, a new theory of belief revision for R, a relevance
logic weaker than RM$_3$, is developed and it is sketched how this new
theory is connected with the AGM theory.
The new theory is also applicable to RM$_3$.
A very general model of belief revision, in which all postulates of the
AGM theory fail, is proposed in~\cite{Pri01a}.
In~\cite{Was11a}, a survey of proposed adaptations of the AGM theory for
non-classical logics, including paraconsistent logics, is provided.

In~\cite{Ari06a,Ari08a,Ari08b}, a general framework is developed for
reasoning where conclusions are drawn according to the most plausible
interpretations,%
\footnote
{In the case of propositional logics, the interpretations are usually
 valuations.
}
i.e.\ the interpretations that are as close as possible to the set of
hypotheses, and also as faithful as possible to the more reliable or
important hypotheses in this set.
This kind of reasoning emerges for example in belief revision with
minimal change and in integration of information from different
autonomous sources.
It also encompasses adaptive reasoning, i.e.\ reasoning where, if a set
of hypotheses can be split up into a consistent part and an inconsistent
part, every assertion that is not related to the inconsistent part and
classically follows from the consistent part, is deduced from the whole
set.
It is very likely that the kind of reasoning covered by the framework
occurs in legal reasoning as well.
The framework focusses on the consequence relations of logics for this
kind of reasoning and it is based on two principles: a distance-based
preference relation on interpretations and prioritized hypotheses.

A well-known paraconsistent logic in which a preference relation on
interpretations underlies the consequence relation is LPm~\cite{Pri91a}
(minimally inconsistent LP).
In~\cite{KM02a}, variants of LP$^{\IImpl,\False,\Both}$ are studied in
which a preference relation on interpretations underlies the consequence
relation.
A characteristic property of logics in which a preference relation on
interpretations underlies the consequence relation is that they are
paraconsistent, even if the point of departure is a non-paraconsistent
logic.
For the sake of completeness, we mention that the above-mentioned
framework generalizes earlier work on preferential reasoning presented
in~\cite{AA98a}.
The work on preferential reasoning presented in~\cite{AL01a} seems to
originate from that earlier work as well.

\section{Concluding remarks}
\label{sect-conclusions}

The discussions of the different paraconsistent logics included in the
survey do not give known theoretical details about them that are not
relevant to a basic understanding of them or their interrelationships.
The details concerned can be found in the cited publications.
The amount of detail that is given in this note differs from one logic
to another for the simple reason that what is known about the questions
concerned differs from one logic to another.

Not all paraconsistent logics discussed in this note have been
investigated in an equally extensive way.
LP$^{\IImpl}$ has been investigated most extensively and D$_2$ has been
investigated least extensively.
However, D$_2$ is mentioned in virtually all publications on
paraconsistent logics.
It seems that the logics C$_n$ are criticized most.
It also seems that, with the exception of paraconsistent annotated
logics, the applications of paraconsistent logics are only looked for
and found in mathematics.

Paraconsistent annotated logics stem from logic programming in the
presence of inconsistent information (see e.g.~\cite{BS89a}).
Other applications of paraconsistent annotated logics include database
query answering~\cite{ABK00a}, negotiation in multi-agent
systems~\cite{HAS05a}, expert system diagnosis~\cite{DRMF06a} and robot
control~\cite{Abe06a} in the presence of inconsistent information.
In some of these applications, it is the case that
$|\tau| = [0,1] \times [0,1]$.
In those cases, the logic concerned has some characteristics of a fuzzy
logic.

Most work on how to choose among the different paraconsistent logics 
focuses on the logical consequence relation of the logics.
In~\cite{BM15b}, properties of the logical equivalence relation turn out
to be of major concern in choosing a paraconsistent logic to build a 
process algebra that allows for dealing with contradictory states on.
The properties of the logical equivalence relation concerned cause an 
extensive reduction of the number of paraconsistent logics to choose 
from (see also~\cite{Mid17a}).
The chosen logic is the logic LP$^{\IImpl,\False}$ mentioned in 
Section~\ref{sect-LPiimpl}.

\bibliographystyle{splncs03}
\bibliography{PCL}

\begin{thebibliography}{10}
\providecommand{\url}[1]{\texttt{#1}}
\providecommand{\urlprefix}{URL }

\bibitem{AN09a}
Abe, J.M., Nakamtsu, K.: A survey of paraconsistent annotated logics and
  applications. International Journal of Reasoning-based Intelligent Systems
  1(1--2),  31--42 (2009)

\bibitem{Abe06a}
Abe, J.M., Torres, C.R., Torres, G.L., Nakamatsu, K., Kondo, M.: Intelligent
  paraconsistent logic controller and autonomous mobile robot {Emmy~II}. In:
  Gabrys, B., Howlett, R.J., Jain, L.C. (eds.) KES 2006, Part II. Lecture Notes
  in Artificial Intelligence, vol. 4252, pp. 851--857. Springer-Verlag (2006)

\bibitem{ANA07a}
Akama, S., Nakamtsu, K., Abe, J.M.: A natural deduction system for annotated
  predicate logic. In: Apolloni, B., Howlett, R.J., Jain, L. (eds.) KES 2007 /
  WIRN 2007, Part II. Lecture Notes in Artificial Intelligence, vol. 4693, pp.
  861--868. Springer-Verlag (2007)

\bibitem{AGM85a}
Alchourr{\'{o}}n, C.E., G{\"{a}}rdenfors, P., Makinson, D.: On the logic of
  theory change: Partial meet contraction and revision functions. The Journal
  of Symbolic Logic  50(2),  510--530 (1985)

\bibitem{AB75a}
Anderson, A.R., Belnap, Jr, N.D.: Entailment, the Logic of Relevance and
  Necessity, vol.~1. Princeton University Press, Princeton NJ (1975)

\bibitem{ABK00a}
Arenas, M., Bertossi, L., Kifer, M.: Applications of annotated predicate
  calculus to querying inconsistent databases. In: Lloyd, J., et~al. (eds.) CL
  2000. Lecture Notes in Artificial Intelligence, vol. 1861, pp. 926--941.
  Springer-Verlag (2000)

\bibitem{Ari06a}
Arieli, O.: Distance-based semantics for multiple-valued logics. Available at
  {\tt http://www2.mta.ac.il/\~{}oarieli/Papers/nmr06.pdf} (2006)

\bibitem{Ari08a}
Arieli, O.: Distance-based paraconsistent logics. International Journal of
  Approximate Reasoning  48(3),  766--783 (2008)

\bibitem{Ari08b}
Arieli, O.: Reasoning with prioritized information by iterative aggregation of
  distance functions. Journal of Applied Logic  6(4),  589--605 (2008)

\bibitem{AA96a}
Arieli, O., Avron, A.: Reasoning with logical bilattices. Journal of Logic,
  Language, and Information  5(1),  25--63 (1996)

\bibitem{AA98a}
Arieli, O., Avron, A.: The value of the four values. Artificial Intelligence
  102(1),  97--141 (1998)

\bibitem{AAZ11b}
Arieli, O., Avron, A., Zamansky, A.: Ideal paraconsistent logics. Studia Logica
   99(1--3),  31--60 (2011)

\bibitem{AAZ11a}
Arieli, O., Avron, A., Zamansky, A.: Maximal and premaximal paraconsistency in
  the framework of three-valued semantics. Studia Logica  97(1),  31--60 (2011)

\bibitem{Arr80a}
Arruda, A.I.: A survey of paraconsistent logic. In: Arruda, A.I., Chuaqui, R.,
  da~Costa, N.C.A. (eds.) Mathematical Logic in Latin America. Studies in Logic
  and the Foundations of Mathematics, vol.~99, pp. 1--41. Elsevier (1980)

\bibitem{Ase66a}
Asenjo, F.G.: A calculus of antinomies. Notre Dame Journal of Formal Logic
  7(1),  103--105 (1966)

\bibitem{Avr91a}
Avron, A.: Natural 3-valued logics --- characterization and proof theory. The
  Journal of Symbolic Logic  56(1),  276--294 (1991)

\bibitem{Avr99a}
Avron, A.: On the expressive power of three-valued and four-valued languages.
  The Journal of Logic and Computation  9(6),  977--994 (1999)

\bibitem{Avr05a}
Avron, A.: Combining classical logic, paraconsistency and relevance. Journal of
  Applied Logic  3(1),  133--160 (2005)

\bibitem{AL01a}
Avron, A., Lev, I.: A formula-preferential base for paraconsistent and
  plausible reasoning systems. Available at {\tt
  http://www.cs.tau.ac.il/\~{}aa/articles/\linebreak[2]plausible-11p.pdf} (2001)

\bibitem{Bat80a}
Batens, D.: Paraconsistent extensional propositional logics. Logique et Analyse
   90--91,  195--234 (1980)

\bibitem{BC04a}
Batens, D., de~Clercq, K.: A rich paraconsistent extension of full positive
  logic. Logique et Analyse  185--188,  220--257 (2004)

\bibitem{BM15b}
Bergstra, J.A., Middelburg, C.A.: Contradiction-tolerant process algebra with
  propositional signals. Fundamenta Informaticae  153(1--2),  29--55 (2017)

\bibitem{BS89a}
Blair, H.A., Subrahmanian, V.S.: Paraconsistent logic programming. Theoretical
  Computer Science  68(2),  135--154 (1989)

\bibitem{CCM07a}
Carnielli, W.A., Coniglio, M.E., Marcos, J.: Logics of formal inconsistency.
  In: Gabbay, D., Guenthner, F. (eds.) Handbook of Philosophical Logic,
  vol.~14, pp. 1--93. Springer-Verlag, Berlin (2007)

\bibitem{Ciu08a}
Ciuciura, J.: Frontiers of the discursive logic. Bulletin of the Section of
  Logic  37(2),  81--92 (2008)

\bibitem{Con07a}
Coniglio, M.E.: Logics of deontic inconsistency. CLE e-Prints  7(4) (2007)

\bibitem{daC74a}
da~Costa, N.C.A.: On the theory of inconsistent formal systems. Notre Dame
  Journal of Formal Logic  15(4),  497--510 (1974)

\bibitem{daC77a}
da~Costa, N.C.A.: A semantical analysis of the calculi {C$_n$}. Notre Dame
  Journal of Formal Logic  18(4),  621--630 (1977)

\bibitem{CB98a}
da~Costa, N.C.A., Bueno, O.: Belief change and inconsistency. Logique \&
  Analyse  41(161--163),  31--56 (1998)

\bibitem{CC86a}
da~Costa, N.C.A., Carnielli, W.A.: On paraconsistent deontic logic. Philosophia
   16(3--4),  293--305 (1986)

\bibitem{dKB07a}
da~Costa, N.C.A., Krause, D., Bueno, O.: Paraconsistent logics and
  paraconsistency. In: Jacquette, D. (ed.) Philosophy of Logic, pp. 791--911.
  Elsevier, Amsterdam (2007)

\bibitem{dSV91a}
da~Costa, N.C.A., Subrahmanian, V.S., Vago, C.: The paraconsistent logic
  {P$\mathcal{T}$}. Zeitschrift f\"{u}r Mathematische Logik und Grundlagen der
  Mathematik  37(9--12),  139--148 (1991)

\bibitem{DRMF06a}
Da~Silva~Filho, J.I., Rocco, A., M{\'{a}}rio, M.C., Ferrara, L.F.: Annotated
  paraconsistent logic applied to an expert system dedicated for supporting in
  an electric power transmission systems re-establishment. In: PSCE 2006. pp.
  2212--2220. IEEE (2006)

\bibitem{DOt85a}
D{\'{}}Ottaviano, I.M.L.: The completeness and compactness of a three-valued
  first-order logic. Revista Colombiana de Matem{\'{a}}ticas  19,  77--94
  (1985)

\bibitem{Dd70a}
D{\'{}}Ottaviano, I.M.L., da~Costa, N.C.A.: Sur un probl\`{e}me de
  {Ja\'{s}kowski}. Comptes Rendus de l\'{}Academie de Sciences de Paris (A--B)
  270,  1349--1353 (1970)

\bibitem{Dun00a}
Dunn, J.M.: Partiality and its dual. Studia Logica  65(1),  5--40 (2000)

\bibitem{DR02a}
Dunn, M., Restall, G.: Relevance logic. In: Gabbay, D., Guenthner, F. (eds.)
  Handbook of Philosophical Logic, vol.~6, pp. 1--128. Kluwer Academic
  Publishers, Dordrecht, second edn. (2002)

\bibitem{GRR08a}
Gabbay, D., Rodrigues, O., Rosso, A.: Belief revision in non-classical logics.
  The Review of Symbolic Logic  1(3),  267--304 (2008)

\bibitem{Gar88a}
G{\"{a}}rdenfors, P.: Knowledge in Flux: Modeling the Dynamics of Epistemic
  States. MIT Press, Cambridge, MA (1988)

\bibitem{HAS05a}
Hasegawa, F.M., {\'{A}}vila, B.C., Shmeil, M.A.H.: A new approach for offer
  evaluation in multi-agent system negotiation based in evidential
  paraconsistent logic. In: Ramos, F.F., Rosillo, V.L., Unger, H. (eds.) ISSADS
  2005. Lecture Notes in Computer Science, vol. 3563, pp. 483--494.
  Springer-Verlag (2005)

\bibitem{Jas99b}
Ja{\'{s}}kowski, S.: On the discussive conjunction in the propositional
  calculus for inconsistent deductive systems. Logic and Logical Philosophy  7,
   57--59 (1999), translated from the Polish edition of 1949

\bibitem{Jas99a}
Ja{\'{s}}kowski, S.: A propositional calculus for inconsistent deductive
  systems. Logic and Logical Philosophy  7,  35--56 (1999), translated from the
  Polish edition of 1948

\bibitem{Kle52a}
Kleene, S.C.: Introduction to Metamathematics. North-Holland, Amsterdam (1952)

\bibitem{KM02a}
Konieczny, S., Marquis, P.: Three-valued logics for inconsistency handling. In:
  Flesca, S., Greco, S., Leone, N., Ianni, G. (eds.) JELIA 2002. Lecture Notes
  in Artificial Intelligence, vol. 2424, pp. 332--344. Springer-Verlag (2002)

\bibitem{Kd79a}
Kotas, J., da~Costa, N.C.A.: A new formulation of discussive logic. Studia
  Logica  38(4),  429--445 (1979)

\bibitem{LL96a}
Lakemeyer, G., Lang, W.: Belief revision in a nonclassical logic. In: G\"{o}rz,
  G., H\"{o}lldobler, S. (eds.) KI-96: Advances in Artificial Intelligence.
  Lecture Notes in Computer Science, vol. 1137, pp. 199--211. Springer-Verlag
  (1996)

\bibitem{Mar02a}
Mares, E.D.: A paraconsistent theory of belief revision. Erkenntnis  56(2),
  229--246 (2002)

\bibitem{McG06a}
McGinnis, C.: Tableau systems for some paraconsistent modal logics. Electronic
  Notes in Theoretical Computer Science  143,  141--157 (2006)

\bibitem{Mid17a}
Middelburg, C.A.: On the strongest three-valued paraconsistent logic contained
  in classical logic. {\tt arXiv:1702.03414 [cs.LO]} (February 2017)

\bibitem{Nel59a}
Nelson, D.: Negation and separation of concepts in constructive systems. In:
  Heyting, A. (ed.) Constructivity in Mathematics. pp. 208--225. North-Holland
  (1959)

\bibitem{Pri79a}
Priest, G.: The logic of paradox. Journal of Philosophical Logic  8(1),
  219--241 (1979)

\bibitem{Pri84a}
Priest, G.: Introduction: Paraconsistent logics. Studia Logica  43(1--2),
  3--16 (1984)

\bibitem{Pri91a}
Priest, G.: Minimally inconsistent {LP}. Studia Logica  50(2),  321--331 (1991)

\bibitem{Pri01a}
Priest, G.: Paraconsistent belief revision. Theoria  67(3),  214--228 (2001)

\bibitem{Pri02a}
Priest, G.: Paraconsistent logic. In: Gabbay, D., Guenthner, F. (eds.) Handbook
  of Philosophical Logic, vol.~6, pp. 287--393. Kluwer Academic Publishers,
  Dordrecht, second edn. (2002)

\bibitem{RS95a}
Restall, G., Slaney, J.: Realistic belief revision. In: Glas, M.D., Pawlak, Z.
  (eds.) WOCFAI '95. pp. 367--378. Angkor (1995)

\bibitem{Sub87a}
Subrahmanian, V.S.: On the semantics of quantitative logic programs. In:
  Heyting, A. (ed.) Fourth IEEE Symposium on Logic Programming. pp. 173--182.
  IEEE Computer Society Press (1987)

\bibitem{Was11a}
Wassermann, R.: On {AGM} for non-classical logics. Journal of Philosophical
  Logic  (2011), {DOI}: 10.1007/s10992-011-9178-2

\end{thebibliography}


\end{document}